\documentclass[11pt]{article}
\usepackage[english]{babel}
\usepackage{graphicx}
\usepackage{setspace}
\usepackage{parskip}
\usepackage{float}
\usepackage{tabularx}
\usepackage{graphicx}
\usepackage{amsmath}
\usepackage{amsthm}
\usepackage{amsfonts}
\usepackage{amssymb}
\usepackage{bm}
\usepackage{titling}
\usepackage[numbers,comma,sort&compress]{natbib}
\usepackage[affil-it]{authblk}
\usepackage{url}
\usepackage[verbose,a4paper,tmargin=2.54cm,bmargin=2.54cm,lmargin=2.54cm,rmargin=2.54cm]{geometry}
\usepackage{lineno}
\usepackage[table]{xcolor}
\usepackage[most]{tcolorbox}
\usepackage[hidelinks]{hyperref} 
\usepackage{booktabs}
\usepackage[page]{appendix}
\usepackage{dsfont}
\usepackage[perpage]{footmisc} 
\usepackage{algorithm}
\usepackage{algpseudocode}
\usepackage{siunitx}
\usepackage{tikz}
\usepackage{titlesec}

\graphicspath{{figures/}}

\titleformat{\section}[block]{\Large\bfseries}{}{0em}{}
\titleformat{\subsection}[block]{\large\bfseries}{}{0em}{}

\title{\bf The impact of strong activity disruption on building energetics\vspace{0.2 in}}

\author[1]{Yu-Hsuan Hsu}
\author[2]{Sara Beery}
\author[3]{Christopher P. Kempes}
\author[3,4]{\\ Mingzhen Lu}
\author[1,3*]{Serguei Saavedra}

\affil[1]{\small Department of Civil and Environmental Engineering, MIT, \protect \\ \small 77 Massachusetts Avenue, Cambridge, MA 02139,  USA \vspace{0.15in}}
\affil[2]{\small Department of Electrical Engineering and Computer Science, MIT, \protect \\ \small 77 Massachusetts Avenue, Cambridge, MA 02139,  USA \vspace{0.15in}}
\affil[3]{\small Santa Fe Institute, \protect \\ \small Santa Fe, NM 87501, USA \vspace{0.15in}}
\affil[4]{\small Department of Environmental Studies, New York University,\protect \\ \small New York, NY 10012, USA \vspace{0.15in}}
\affil[*]{\small To whom correspondence should be addressed. \protect \\ E-mail: sersaa@mit.edu}
\date{}

\begin{document}
\maketitle

\clearpage

\setstretch{1.5}

\section{Abstract}

Evidence shows that biological organisms tend to be more energetically efficient per unit size. These scaling patterns observed in biological organisms have also been observed in the energetic requirements of cities. However, at lower levels of organization where energetic interventions can be more manageable, such as buildings, this analysis has remained more elusive due to the difficulties in collecting fine-grained data. Here, we use the maintenance energy usage in buildings at the Massachusetts Institute of Technology (MIT) from 2009 to 2024 to analyze energetic trends at the scale of individual buildings and their sensitivity to strong external perturbations. We find that, similar to the baseline metabolism of biological organisms, large buildings are on average $24\%$ more energetically efficient per unit size than smaller buildings. Because it has become debatable how to better measure the efficiency of buildings, this scaling pattern naturally establishes a baseline efficiency for buildings, where deviations from the mean would imply a more or less efficient building than the baseline according to volume. This relative efficiency progressively increased to $34\%$ until 2020. However, the strong activity disruption caused by the COVID-19 pandemic acted as a major shock, removing this trend and leading to a reversal to the expected 24\% baseline level. This suggests that energetic adaptations are contingent on relatively stable conditions.

\noindent  {\textbf{Keywords}: energy, metabolism, scaling, adaptation, society}

\clearpage



\section{Introduction}

Efforts to address global warming have primarily focused on renewable energy innovation and carbon pricing mechanisms \cite{REN21_2024, WorldBank2024}, while reducing building energy use has received less attention. Yet, the building sector plays a crucial role in global energy use and carbon emissions, accounting for 21\% of global GHG emissions (12 $GtCO_2$-eq) and 31\% of global final energy demand (128.8 EJ) in 2019 \cite{Cabeza2022}. Notably, the global floor area of buildings in cities is projected to increase from 244 billion m$^2$ in 2020 to 427 billion m$^2$ by 2050 \cite{IEA2022}. This means that the world is expected to add about 183 billion m$^2$ of new floor area to the global building stock, the equivalent of adding two entire Boston cities every month for 30 years. It becomes imperative then to understand the physical and socio-economic factors driving building energetics for a sustainable future.

The study of energetics has a long tradition in the life sciences \cite{Boltzmann1905,lotka1922contribution,Schrodinger}. Evidence shows that biological organisms tend to be more energetically efficient per unit size---known as  Kleiber’s Law \cite{Kleiber}. Typically, the dependence of metabolism ($y$) on body mass ($x$) takes the nonlinear form $y=\beta x^{\alpha}$, where $\beta$ is a taxon-specific normalization constant and $\alpha$ represents the scaling exponent \cite{West}. For example, in the animal kingdom, the scaling exponent is typically observed to be sublinear around $\alpha=3/4$ \cite{delong2010shifts}. This implies that every unit increase in size, there is an average $25\%$ decrease in energy requirements per unit size. Put simply, while an elephant requires more energy than a mouse; the elephant can be sustained more efficiently than the mouse. Interestingly, the scaling exponent seems to be different across the tree of life. It has been shown that ordering organisms from relatively simple (e.g., prokaryotes) to relatively complex (e.g., metazoans), the scaling exponent changes from superlinear ($\alpha>1$) to sublinear ($\alpha<1$) \cite{delong2010shifts}. These scaling changes are expected to arise due to adaptation processes driven by body plan innovations filtered by natural selection \cite{delong2010shifts,Kempes,West}.

The scaling patterns observed in biological organisms have also been observed in the energetic requirements of cities \cite{BETTENCOURT,Ayres}. Requirements linked to infrastructure tend to depend sublinearly on cities' area, and in contrast requirements linked to socio-economic factors tend to depend superlinearly on cities' area \cite{BETTENCOURT}. Importantly, it has been shown that such scaling exponents in cities can change relatively fast \cite{Lu}, reflecting the differences in speed between biological and socio-economic adaptations \cite{innovations}. However, at lower levels of organization where energetic interventions can be more manageable, such as buildings, this analysis has remained more elusive due to the difficulties in collecting fine-grained data \cite{Smil}. Indeed, a building as a physical individual object does not undergo significant evolution (though one can argue that modern retrofit and improvements all count as evolution, in fact, a lot of iconic buildings cost more and more for modern improvements nowadays), but a building as a human-made system has gone through human selection since early settlement, from the makeshift tents in Stone Age (million years ago- $\sim 10,000$ year ago) to the modern day architecture. Moreover, it remains unclear the potential impact of strong activity disruptions on building energetics. To address this gap, we use the maintenance energy usage in buildings at the Massachusetts Institute of Technology (MIT) from 2009 to 2024 to analyze energetic trends at the scale of individual buildings\cite{mitdata}. This usage measurement entails the total of both electricity and gas, no information was available for energy use through materials. In particular, we investigate the relationship between energy use and building size, the dynamics of the scaling exponents, and the impact of activity disruptions caused by COVID-19. Lastly, we discuss the implications of our work for sustainable city design.

\section{Results}

Volume is one of the most distinctive characteristics of buildings. As expected, studies have found a strong association between building volume (area and height) and energy and gas usage \cite{Malakoutian,Godoy-Shimizu}. This relationship is mainly due to factors such as air-conditioning and heating demands driven by exposure to environmental conditions. Similar to other buildings in cities, MIT buildings have diverse usages, ranging from labs, offices, mixed, to residencies (Fig. 1a). Because we only have information about floor area and maximum height, we estimate volume assuming buildings are rectangular prisms. This assumption, while course, does not affect our conclusions as this volume is fixed across years. MIT buildings range from relatively small ($3,552$ m$^3$ equivalent to 1.42 Olympic Swimming Pools, building 51 Fig. 1a) to relatively large ($1,850,253$ m$^3$ equivalent to 1.85 Empire State Buildings, building 32 Fig. 1a), whereas the median volume is $183,918$ m$^3$.

Focusing our analysis on the energetics of 86 buildings reported in 2024, we found that energy use has a scaling relationship with volume defined by an exponent of $\alpha=0.76$ (Fig. 2a) and constant $\beta=0.15$ (GJ per unit volume) (Fig. 2b). This reveals that, similar to the baseline metabolism of biological organisms, large buildings are on average $\sim 24\%$ more energetically efficient per unit size than small buildings. Because it has become debatable how to better measure the efficiency of buildings \cite{Smil}, this scaling pattern naturally establishes a baseline efficiency for buildings, where deviations from the mean (i.e., positive or negative regression residuals) would imply a more or less efficient building than the baseline according to volume (Fig. 1a shows standardized residuals).

From 2009 to 2024, MIT (hosting almost 30,000 people \cite{mitdata}) had an average energy use in buildings of $2.6$ PJ (a maximum of 2.88 PJ in 2011 and a minimum of 2.5 PJ in 2024). This usage is equivalent to 69,000 average U.S. households. However, changes in energy use across years were not proportional for every building. Figure 2a shows that the scaling exponent progressively decreased from $\alpha=0.73$ in 2009 to $\alpha=0.66$ in 2020 (Fig. 2b). In other words, the relative efficiency of large buildings increase to $34\%$ (from the $24\%$ baseline). In contrast, the normalization constant $\beta$ progressively increased from $\beta=0.46$ in 2009 to $\beta=0.68$ in 2020. That is, the demand for energy per unit volume increased, reflecting a proportionally larger usage of smaller buildings. Note that these statistical values are typically considered meaningful in the area of biology and social sciences (given the noise or scarcity of the data) \cite{Lu}. In sum, these results show that efficiency improvements exhibited by MIT until 2020, were coming precisely from the most energy-demanding buildings.

However, the strong activity disruption caused by the COVID-19 pandemic led to a reversal to the mean from 2021-2024, bringing efficiency back to the baseline of $24\%$ relative efficiency (Fig. 3a-b). The reversal to the mean effect implies that values far from the expectation in the present tend to be followed by values closer to the expectation in the future \cite{Barnett,Samuels}. To test this effect, we calculated the correlation between residuals in year 2009 ($\mathbf{r_0}$) and subsequent residuals ($\mathbf{r_t}$). This correlation implies the linear relationship $\mathbf{z_t}=\rho \cdot \mathbf{z_0}$, where $\mathbf{z_t}=\mathbf{r_t}\cdot \sigma_{\mathbf{r_t}}^{-1}$ (respectively, $\mathbf{z_0}=\mathbf{r_0}\cdot \sigma_{\mathbf{r_0}}^{-1}$) represents the vector of standardized residuals (Fig. 1a) and $\rho$ corresponds to the Pearson correlation. The weaker the correlation (and statistically detectable), the stronger the reversal to the mean \cite{Barnett}. In line with this concept, Figure 3a shows that correlations were weaker from 2021-2023. Additionally, the reversal to the mean effect requires that the conditional expected value of observations in the future are bounded by the conditional expected value of observations in the present \cite{Samuels}. Using again standardized residuals as observations, the previous statement formally implies $\mu=0 \leq E[\mathbf{z_t}|\mathbf{z_0}>h]<E[\mathbf{z_0}|\mathbf{z_0}>h]$ and $\mu=0\geq E[\mathbf{z_t}|\mathbf{z_0}<h]>E[\mathbf{z_{0}}|\mathbf{z_0}<h]$, where $h$ is a standardized threshold value (interpreted as standard deviations). Figure 3b shows the strongest reversal to the mean ($\mu=0$) within the period 2021-2024 across different threshold values.

\section{Discussion}

The findings of this study reveal a distinct scaling relationship between building energy use and volume, mirroring patterns observed in biological metabolism \cite{BETTENCOURT,West}. This discovery aligns with well-established metabolic scaling relationships, where larger organisms exhibit greater efficiency in energy usage per unit size (Kleiber’s Law \cite{Kleiber}). Specifically, the energy use of buildings follows a sublinear scaling as a functio of size, indicating that larger buildings are inherently more energy-efficient per unit volume compared to smaller ones. While we did not have information about people in each building (due to the complexity in the day-to-day dynamics at MIT), it is expected that the number of people will grow proportionally to building size. Notably, the decrease in energy use per unit volume in larger buildings likely stems from structural and operational factors, such as centralized circulation systems (less energy dissipation) and economies of scale in energy distribution. Future work should further investigate the extent to which human-made systems can indeed have similar energetic developments as biological organisms \cite{BETTENCOURT}.

Over the years preceding 2020, the relative efficiency advantage of large buildings progressively increased, reflecting an adaptation likely influenced by advancements in energy management and sustainability practices. However, the strong activity disruptions caused by the COVID-19 pandemic abruptly altered this trajectory, leading to a reversal to baseline 24\% efficiency levels (Fig. 2b). This alteration may be the consequence of eliminating adaptive practices reducing energy waste over the years. The rapid shift in energy efficiency trends highlights the sensitivity of building energetics to external shocks. This reversal to the mean suggests that, while efficiency gains can be achieved through sustainability-driven targets, they remain contingent on stable usage patterns and adaptive management strategies \cite{cities_covid}. In general, this resonates with biological innovations (or adaptations) that tend to occur during periods of relatively stable environments \cite{Kempes,delong2010shifts}. Yet, under strong perturbations, energetic processes may reverse to the expectations set by first principles.

These insights carry significant implications for sustainable building design and energy policy. If larger buildings naturally exhibit greater efficiency per unit volume, then urban planning could expand the construction of high-volume structures to increase energy performance and reduce floor area. This may also help in reducing land-use changes with lower ecological impact. Yet, in absolute terms, larger buildings can require excessive energy flows that may not be met in specific areas or may rely too heavily on fossil fuels. Therefore, optimization process can take into account efficiency, absolute usage, area, carbon footprint, and number of people served or affected. Moreover, future policies should acknowledge that energy efficiency adaptations may be fragile to external disruptions and promote sustainable practices in both new and existing buildings to ensure long-term memory. This can be achieved by implementing adaptive energy systems, such as smart grid integration \cite{Ardebili2024}, which help buildings dynamically respond to fluctuations and sustain efficiency gains even during unexpected challenges.

Despite several guidelines being proposed, this study presents certain limitations that warrant further exploration. The analysis is confined to a single institution, and while MIT provides a rich dataset due to its diverse building types and functions, expanding this research to other universities or urban environments would help validate the applicability of these findings. Additionally, the study does not account for energy usage variations within individual buildings, such as differences in energy use patterns among laboratories, offices, residential halls, and mixed-use buildings. Future research should incorporate finer-grained data, such as number of people, to better understand the micro-level determinants of energy efficiency within buildings. Understanding these dynamics provides valuable guidelines for designing more resilient and energy-efficient buildings, particularly in the context of global sustainability challenges. Continued investigations should focus on delving deeper into the underlying mechanisms of energy scaling, enabling more effective policies and innovations in building energy management.

\vspace{0.1 in}

\subsection*{Abbreviations}

MIT: Massachusetts Institute of Technology

GHG: Green House Gas

US: United States of America

\subsection*{Declarations}

\noindent {\bf Ethics approval and consent to participate}
This study does not incorporate any personal information.

\noindent {\bf Consent for publication}. All of the authors consent to publication.

\noindent  {\bf Availability of data and material}: The code and data associated with this work can be accessed at \url{https://github.com/hsuanmina/building-energetics}. The code and data will be deposited at Zenodo upon acceptance.

\noindent {\bf Competing interests.} The authors declare no competing interests.

\noindent  {\bf Funding.} SS acknowledges support from the National Science Foundation under Grant No. DEB-2436069 and MIT Google Program for Computing Innovation. SB acknowlesges support from MIT-IBM Watson AI Lab, MIT J-WAFS, the NSF/NSERC ABC Global Center on AI and Biodiversity Change, and the Schmidt AI2050 Fellowship.

\noindent  {\bf Author contributions}: SS designed and supervised the work. YHH performed the work. All authors analyzed results. SS and YHH wrote a first version of the manuscript. All authors contributed to the revision of the manuscript.

\noindent  {\bf Acknowledgments.} We are grateful to Yu Cheng for guidance on data, and Timothy Gutowsky, Anette (Peko) Hosoi, Julie Newman, Washington Taylor, Maria Yang, and Vicky Yang for insightful discussion that led to the development of this work.

\clearpage

\begin{figure}[ht]
\centering
\includegraphics[width = \textwidth]{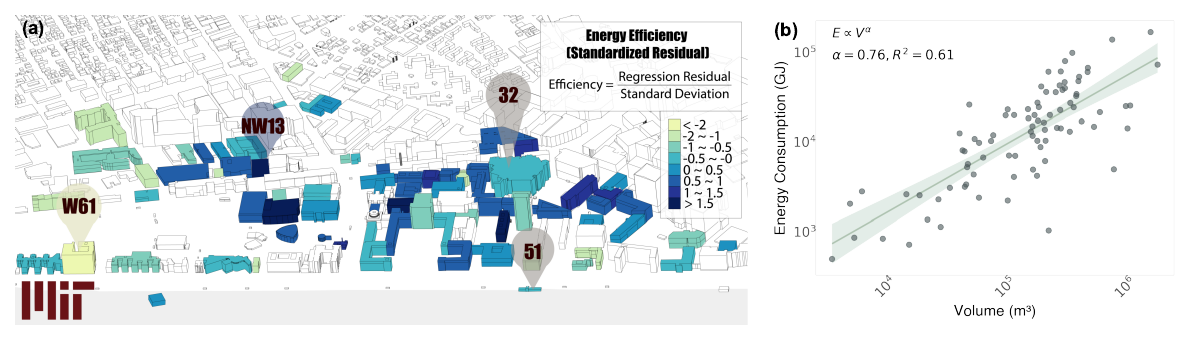}
\caption{
    \textbf{Building energetics at MIT in 2024.} \textbf{(a)} Map of MIT buildings illustrating the volume range. Buildings 32 and 51 are shaded in gray representing the largest and smallest, respectively. The lighter the color, the higher the energy efficiency of a building relative to its volume. Buildings in the lightest and darkest two shades are specifically highlighted to represent the extremes of relative efficiency (among the buildings analyzed, Building W61 demonstrates the highest relative efficiency, while Building NW13 exhibits the lowest relative efficiency). Formally, relative efficiency is measured by standardized residuals ($\mathbf{z}=\mathbf{r}\cdot \sigma_{\mathbf{r}}^{-1}$, see main text), considering that the expectation is given by the scaling relationship $\text{energy} \propto \text{volume}^{\alpha = 0.76}$. This scaling relationship is similar by the one exhibited by biological organisms. \textbf{(b)} Scaling relationship between energy use and volume. Each point corresponds to 1 out of 86 different buildings, the line represents the expectation (used in panel \textbf{a} to calculate standardized residuals), and the shaded area corresponds to the $95 \%$ confidence interval (also used in panel \textbf{a} to calculate standardized residuals).
    }
    \label{fig:fig1}
\end{figure}

\clearpage

\begin{figure}[ht]
\centering
\includegraphics[width = \textwidth]{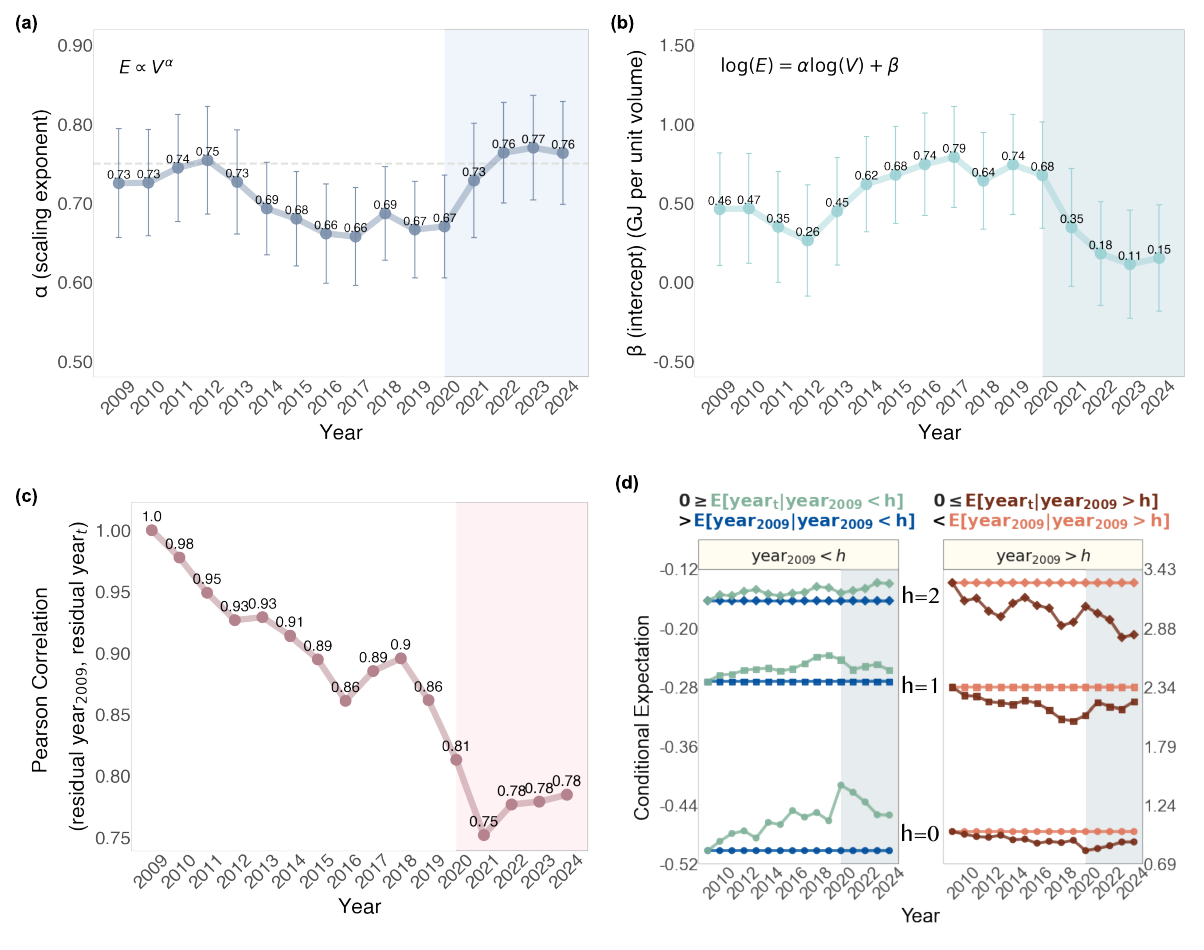}
\caption{
    \textbf{Building energetics at MIT before and after the COVID-19 pandemic.} \textbf{(a)} The scaling exponent ($\alpha$) defining the dependence of energy use on volume changed across years. At the begging of the time series, the scaling exponent is on average close to $\alpha=0.75$. This scaling exponent decreased to $\alpha=0.66$ before 2020. However, during the period after COVID-19 (shaded area), the scaling exponent return to baseline levels ($\alpha=0.75$). \textbf{(b)} The normalization constant ($\beta$) defining the energy use per unit volume changed across years, reflecting that smaller buildings started to demand more energy.
    }
    \label{fig:fig2}
\end{figure}

\clearpage

\begin{figure}[ht]
\centering
\includegraphics[width = \textwidth]{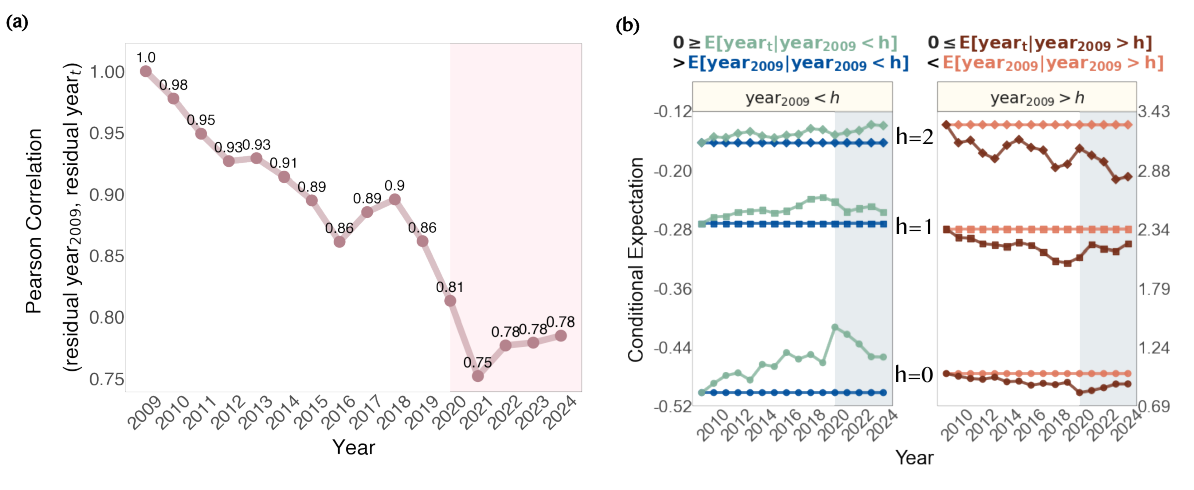}
\caption{
    \textbf{Reversal to the mean.} \textbf{(a)} Pearson correlation ($\rho$) between residuals in year 2009 and year $t$ ($\mathbf{z_t}=\rho \cdot \mathbf{z_{2009}}$). \textbf{(b)} Conditional expected value of observations in the year $t$ ($E[\mathbf{z_t}|\mathbf{z_{2009}}>h]$ and $E[\mathbf{z_t}|\mathbf{z_{2009}}<h]$) are bounded by the conditional expected value of observations in year 2009 ($E[\mathbf{z_{2009}}|\mathbf{z_{2009}}>h]$ and $E[\mathbf{z_{2009}}|\mathbf{z_{2009}}<h]$), where $h$ is a standardized threshold value. The weaker the correlation and the closer to the mean ($\mu=0$), the stronger the reversal to the mean (shaded areas).
    }
    \label{fig:fig2}
\end{figure}

\clearpage

\bibliographystyle{pnas}
\bibliography{ref}

\end{document}